\documentclass[prd,showpacs,showkeys,twocolumn]{revtex4}
\usepackage{amsmath}
\usepackage{lipsum}
\usepackage{amssymb}
\usepackage{latexsym}
\usepackage{graphics}
\usepackage{eurosym}
\usepackage{color,soul}
\usepackage{amsfonts}
\usepackage{graphicx}
\usepackage[font={footnotesize,it}]{caption}
\usepackage{calrsfs}

\def\P{{\cal{P}}}

\def\eqref#1{{(\ref{#1})}}

\def\be{\begin{equation}}
\def\ee{\end{equation}}
\def\beq{\begin{eqnarray}}
\def\eeq{\end{eqnarray}}

\begin{document}

\title{Charged Thin-shell Gravastars in  Noncommutative Geometry}

\author{Ali \"{O}vg\"{u}n}
\email{ali.ovgun@pucv.cl}

\affiliation{Instituto de F\'{\i}sica, Pontificia Universidad Cat\'olica de
Valpara\'{\i}so, Casilla 4950, Valpara\'{\i}so, Chile}
\affiliation{Physics Department, Eastern Mediterranean University, Famagusta,
Northern Cyprus, Turkey}

\author{Ayan Banerjee}
\email{ayan_7575@yahoo.co.in}
\affiliation{Department of Mathematics, Jadavpur University, Kolkata 700 032,
West Bengal, India}

\author{Kimet Jusufi}
\email{kimet.jusufi@unite.edu.mk}
\affiliation{Physics Department, State University of Tetovo, Ilinden Street nn, 1200, Tetovo,
Macedonia}
\affiliation{Institute of Physics, Faculty of Natural Sciences and Mathematics, Ss. Cyril and Methodius University, Arhimedova 3, 1000 Skopje, Macedonia}

\date{\today }
\begin{abstract}
In this paper we construct a charged thin-shell gravastar model within the context of noncommutative geometry. To do so, we choose the interior of the nonsingular de Sitter spacetime with an exterior charged noncommutative solution by cut-and-paste technique  and apply the generalized junction conditions. We then investigate the stability of a charged thin-shell gravastar under linear perturbations around the static equilibrium solutions as well as the thermodynamical stability of the charged gravastar. We find the stability regions by choosing appropriate parameter values which is located sufficiently close to the event horizon.


\end{abstract}

\maketitle

\section{Introduction}
One of the most interesting and challenging problems in modern astrophysics is related to compact astrophysical objects like black hole which is widely accepted. The black holes are the end-point of a complete gravitational collapse of the massive star, that can described by the Einstein theory of gravity contains singularities and surrounded by a boundary from which nothing, not even light, can escape. The event horizon of a black hole which acts like a one-way membrane, is a boundary between its exterior and its interior spacetime. Astronomers have found convincing evidence for the existence of supermassive black hole, especially the one corresponding to SgrA* in the Milky Way \cite{Guillessen} has established the concept of a black hole. However, extending the concept of Bose-Einstein condensate \cite{Mielke} to gravitational systems, gravitational vacuum star (gravastar) was proposed as an alternative to black holes by Mazur and Mottola (MM) \cite{Mazur}, which do not involve horizons and could be stabilized under the exotic states of matter. In this purpose, they use the famous cut and paste technique with Israel junction conditions \cite{Israel}. There are many applications of this method of cut and paste technique such as thin-shell wormholes  \cite{Bhar,Musgrave:1995ka,Jusufi:2016eav,Ovgun:2016ujt,Ovgun:2016ijz,Ovgun:2015una,Halilsoy:2013iza}.

In this model it a multi layered structure has been introduced: a de Sitter geometry in the interior filled with constant positive (dark) energy density  accompanied by isotropic negative pressure $p = -\rho$, while the exterior is defined by a Schwarzschild geometry, separated by a thin shell of stiff matter implying that the configuration of a gravastar. Moreover, the gravastar model has no singularity at the origin and no event horizon \cite{Pani:2009hk,Uchikata:2016qku,Kubo:2016ada,Lobo:2015lbc,Pani:2015tga,Sakai:2014pga,Pani:2012zz,MartinMoruno:2011rm,Pani:2010em,Pani:2009ss,Rocha,Matos,Rocha1}.

 Therefore, these alternative models is quite fascinating because it could solve two fundamental problems, one is singularity problem and the other is information loss paradox which are associated with black holes solutions. After this new emerging picture several researchers have analysed the gravastar solutions using different approaches.  A different development of the thick shell anisotropic gravastar model  idea has been developed by  Cattoen et al. \cite{Carter},  with continuous profiles for the energy density and the anisotropic pressures. One development of the gravastar idea went in the direction of  stability analysis against radial perturbations by Visser and Wiltshire \cite{Visser}, with phase transition layer was replaced by a single spherical $\delta$ -shell. These facts frequently motivated other possibilities for the interior
solution have been considered. Among them  Bili$\acute{c}$ \emph{et al}. \cite{Bilic} have replaced the de-Sitter
interior by a Born-Infeld phantom. Recently, the the gravastar solution extended by introducing an electrically charged component in \cite{Horvat} and charged gravastar admitting conformal motion has proposed in \cite{Usmani}.
Further expanding the work Banerjee \emph{et al}. have propose the braneworld gravastar configuration which is alternative to braneworld black hole \cite{Banerjee76}. This theoretical prediction is strongly supported by the
different authors and for more comprehensive review is provided in \cite{Cecilia}.

The main topic that we would like to address in this paper is the
finding of exact charged thin-shell gravastar solutions in the context of noncommutative geometry
where coordinates of the target spacetime become \emph{noncommutating} operators on a D-brane \cite{Witten} as :
$[\hat{x}^{\mu}, \hat{x}^{\nu}]$ = $i\vartheta^{\mu\nu}$, where $\hat{x}$ and $i\vartheta^{\mu\nu}$ are the coordinate
operators and an antisymmetric tensor of dimension (length)$^2$, which determines the fundamental cell discretization
of spacetime. In addition to noncommutativity eliminates is characterized by a Gaussian function distribution
with a minimal width $\sqrt{\theta}$, i.e. a smeared particle, instead of the Dirac-delta function distribution.
In spite of the progress a lot of work have been done on black holes with such Gaussian sources so far 
like higher dimensional black hole \cite{Rizzo}, charged black hole solutions \cite{Ansoldi} and charged rotating black hole solution \cite{Smailagic(2010)}. A way of implementing the energy density of a static and spherically symmetric, smeared and particle-like gravitational source has been considered in the following form \cite{Spallucci:2009zz}  :
\begin{equation}
\rho_{\theta}=\frac{M}{(4\pi\theta)^{\frac{3}{2}}}e^{-\frac{r^{2}}{4\theta}},
\end{equation}
where the mass M is diffused throughout a region of linear dimension due $\sqrt{\theta}$
to the uncertainty. 

Recently, one consider that the LIGO detectors measure the first direct signal of the gravitational wave from rotating gravastars comparing the real and imaginary parts of the ringdown signal of GW150914 and they concluded that the modeling of the ringdown of GW150914 from the rotating gravastar is not possible \cite{ligo}.

Further research on noncommutative geometry the most significant development has been
performed for obtaining an exact solutions of Self-sustained traversable wormholes \cite{Garattini},   
thin-Shell wormholes \cite{Pete} and gravastar solutions in higher and lower dimensional spacetime \cite{Banerjee1}
etc. The main topic that we would like to address in this paper is that to  
find an exact gravastar solutions in the context of noncommutative geometry,
and explore their physically accepted properties. The plan of our paper is organized as follows. 
In Sec. II we construct the generic structure equations of charged gravastars,
in the context of noncommutative geometry and specifying the mass function. In Sec. III
we discuss the matching conditions at the junction interface and determine the surface stresses.
In Sec. IV we investigate the stability of the charged thin-shell gravastar. In Sec. V we shall we consider the thermodunamical stability.  Finally, in Sec. VI, we comment on our results.

\section{Exterior Of Gravastars: Noncommutative geometry inspired Charged BHs}
The metric of a noncommutative charged black hole is described by the
metric given by \cite{Ansoldi},
\begin{equation}
ds^{2}=-f(r)dt^{2}+f(r)^{-1}dr^{2}+r^{2}d\Omega^{2},
\end{equation}
\\
with $ f(r)=\left(1-\frac{2M_{\theta}}{r}+\frac{Q_{\theta}^{2}}{r^{2}}\right)$, 
where the mass and charge functions are defined by
\begin{equation} M_{\theta}(r)=\frac{2M}{\sqrt{\pi}}\gamma\left(\frac{3}{2},\frac{r^{2}}{4\theta}\right), \end{equation}
\begin{equation}Q_{\theta}(r)=\frac{Q}{\sqrt{\pi}}\sqrt{\gamma^{2}\left(\frac{1}{2},
\frac{r^{2}}{4\theta}\right)-\frac{r}{\sqrt{2\theta}}\gamma\left(\frac{1}{2},\frac{r^{2}}{2\theta}\right),} \end{equation}
and
\begin{equation}
\gamma\left(\frac{a}{b},x\right)=\int_{0}^{x}u^{\frac{a}{b}-1}e^{-u}du.
\end{equation}

\begin{widetext}
Here, the metric (2) lead to the result 
\begin{equation}
f(r)=1-\frac{4M}{r\sqrt{\pi}}\gamma\left(\frac{3}{2},\frac{r^{2}}{4\theta}\right)+\frac{Q^{2}}{r^{2}\pi}\left[\gamma^{2}\left(\frac{1}{2},\frac{r^{2}}{4\theta}\right)-\frac{r}{\sqrt{2\theta}}\gamma\left(\frac{1}{2},\frac{r^{2}}{2\theta}\right)\right]
\end{equation}
where M is the total (constant) mass of the system and for the commutative case when 
r/$\sqrt{\theta} \rightarrow \infty$, the smeared-like mass descends to the point-like
mass, i.e. $M_{\theta} \rightarrow M$. It is also that $Q$ is the total charge of the black hole. It
is noted that for large $r$, Reissner-Nordström black hole will be obtained. 
The horizon radius ($r_{h}$) can be found where $f(r_{h})=0$ in
other words.
\end{widetext}

\section{STRUCTURE EQUATIONS OF CHARGED GRAVASTARS}
To construct the charged gravastars, first we consider two noncommutative geometry inspired charged spacetime manifolds. The exterior is defined by ${M_{+}}$, and the interior is ${ M_{-}}$. Then we join them together by using the cut and paste method across a surface layer $\Sigma$ \cite{Lobo:2015lbc}. 
The metrics of interior is the nonsingular de Sitter spacetimes:

\begin{equation}
ds^{2}=-(1-\frac{r_{-}^{2}}{\alpha^{2}})dt_{-}^{2}+(1-\frac{r_{-}^{2}}{\alpha^{2}})^{-1}dr_{-}^{2}+r_{-}^{2}d\Omega_{-}^{2}
\end{equation}
and exterior of noncommutative geometry inspired charged spacetimes:
\begin{equation}
ds^{2}=-f(r)_{+}dt_{+}^{2}+f(r)_{+}^{-1}dr_{+}^{2}+r_{+}^{2}d\Omega_{+}^{2}
\end{equation}
with 
\begin{equation} 
f(r)_{+}=\left(1-\frac{2M_{\theta+}}{r}+\frac{Q_{\theta+}^{2}}{r^{2}}\right).
\end{equation} 
Note that $\pm$ stands for the exterior and interior geometry, respectively.

The induced metrics  are $g_{ij}^{+}$
and $g_{ij}^{-}$, respectively. It is assumpted that $g_{ij}^{+}(\xi)=g_{ij}^{-}(\xi)=g_{ij}(\xi)$,
where the hypersurface coordinates $\xi^{i}=(\tau,\theta,\phi)$. Our aim is to glue${ M_{+}}$
and ${ M_{-}}$ at their boundaries to obtain a single manifold ${M}$ so that ${ M}={M_{+}}\cup{ M_{-}}$,
at the boundaries $\Sigma=\Sigma_{+}=\Sigma_{-}$.

To calculate the stress-energy tensor components, we use the intrinsic metric on $\Sigma$ as follows:
\begin{equation} 
ds_{\Sigma}^{2}=-d\tau^{2}+a(\tau)^{2}\,(d\theta^{2}+\sin^{2}{\theta}\,d\phi^{2}).
\end{equation} 
Then we use the Einstein field equation, $G_{{\mu}{\nu}}=8\pi\,T_{{\mu}{\nu}}$, here it is noted that $c=G=1$. Note that the junction surface is located at $x^{\mu}(\tau,\theta,\phi)=(t(\tau),a(\tau),\theta,\phi)$. 
One finds the unit normal vectors respect to the junction surface are following \cite{Lobo:2015lbc}: 
 \begin{widetext}
\begin{equation} 
n_{-}^{\mu}=\left(\frac{1}{\left(1-\frac{a^{2}}{\alpha^{2}}\right)}\dot{a},\sqrt{\left(1-\frac{a^{2}}{\alpha^{2}}\right)+\dot{a}^{2}},0,0\right)\,,
\end{equation} 
\begin{eqnarray}
n_{+}^{\mu}=\left(\frac{1}{1-\frac{2M_{\theta+}}{a}+\frac{Q_{\theta+}^{2}}{a^{2}}}\dot{a},\sqrt{1-\frac{2M_{\theta_{+}}}{a}+\frac{Q_{\theta+}^{2}}{a^{2}}+\dot{a}^{2}},0,0\right)\,.\label{normal}
\end{eqnarray}
 where the overdot stands for a derivative with respect to $\tau$.
 For the spherical symmetric spacetimes, the condition of the normal vectors is
$n^{\mu}n_{\mu}=+1$. The extrinsic curvatures are calculated by the following equation \cite{MartinMoruno:2011rm}:
\begin{eqnarray}
K_{ij}^{\pm}=-n_{\mu}\left(\frac{\partial^{2}x^{\mu}}{\partial\xi^{i}\,\partial\xi^{j}}+\Gamma_{\;\;\alpha\beta}^{\mu\pm}\;\frac{\partial x^{\alpha}}{\partial\xi^{i}}\,\frac{\partial x^{\beta}}{\partial\xi^{j}}\right)\,.\label{extrinsiccurv}
\end{eqnarray}
\end{widetext}
so it is found as follows:
\begin{eqnarray}
K_{\;\;\theta}^{\theta\;-} & = & \frac{1}{a}\,\sqrt{\left(1-\frac{a^{2}}{\alpha^{2}}\right)+\dot{a}^{2}}\;,\label{genKplustheta-1}\\
K_{\;\;\tau}^{\tau\;-} & = & \left\{ \frac{\left(\ddot{a}-\frac{a}{\alpha^{2}}\right)}{\sqrt{\left(1-\frac{a^{2}}{\alpha^{2}}\right)+\dot{a}^{2}}}\right\} \,,\label{genKminustautau-1}
\end{eqnarray}
\begin{eqnarray}
K_{\;\;\theta}^{\theta\;+} & = & \frac{1}{a}\,\sqrt{1-\frac{2M_{\theta+}}{a}+\frac{Q_{\theta+}^{2}}{a^{2}}+\dot{a}^{2}}\;,\label{genKplustheta}\\
K_{\;\;\tau}^{\tau\;+} & = & \left\{ \frac{\ddot{a}+\frac{(M_{\theta+}a)-Q_{\theta+}^{2}}{a^{3}}}{\sqrt{1-\frac{2M_{\theta+}}{a}+\frac{Q_{\theta+}^{2}}{a^{2}}+\dot{a}^{2}}}\right\} \,,\label{genKminustautau}
\end{eqnarray}

It is noted that the prime is for a derivative with respect to the $a$. Then we calculate the discontinuity as follows: $\kappa_{ij}=K_{ij}^{+}-K_{ij}^{-}$.

The stress-energy tensors $S_{\;j}^{i}$ on $\Sigma$ are calculated by following:
\begin{equation}
S_{\;j}^{i}=-\frac{1}{8\pi}\,\left(\kappa_{\;j}^{i}-\delta_{\;j}^{i}\;\kappa_{\;k}^{k}\right)\,.
\end{equation}
Then using the relation of $S_{\;j}^{i}={\rm diag}(-\sigma,\P,\P)$, one can find the surface energy density, $\sigma$,
and the surface pressure, $\P$, as follows \cite{Lobo:2015lbc}:

\begin{figure}[h!]
\includegraphics[width=0.45\textwidth]{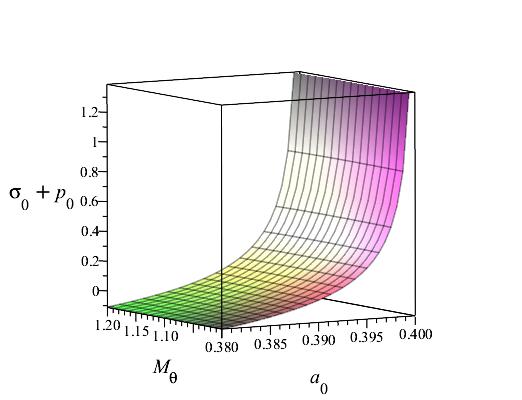} %
\caption{\small \textit{We plot $\sigma_0+p_0$ as a function of $M_{\theta}$ and $a_{0}$. We choose $Q_{\theta}=1$ and $\alpha=0.4$. Note that in this region the NEC is satisfied.} }
\end{figure}

\begin{widetext}
\begin{eqnarray}
\sigma & =-\frac{\kappa_{\;\theta}^{\theta}}{4\pi}= & -\frac{1}{4\pi a}\left[\sqrt{1-\frac{2M_{\theta+}}{a}+\frac{Q_{\theta+}^{2}}{a^{2}}+\dot{a}^{2}}\;-\sqrt{\left(1-\frac{a^{2}}{\alpha^{2}}\right)+\dot{a}^{2}}\;\right],\label{gen-surfenergy2}\\
{\cal P} & =\frac{\kappa_{\;\tau}^{\tau}+\kappa_{\;\theta}^{\theta}}{8\pi}= & \frac{1}{8\pi a}\left[\frac{1+\dot{a}^{2}+a\ddot{a}-\frac{M_{\theta+}}{a}}{\sqrt{1-\frac{2M_{\theta+}}{a}+\frac{Q_{\theta+}^{2}}{a^{2}}+\dot{a}^{2}}}-\frac{\left(1+a\ddot{a}+\dot{a}^2-\frac{2a^2}{\alpha^{2}}\right)}{\sqrt{\left(1-\frac{a^{2}}{\alpha^{2}}\right)+\dot{a}^{2}}}\right].\nonumber \\
\label{gen-surfpressure2}
\end{eqnarray}
Then it is found as follows: 
\begin{eqnarray}
\sigma+2\P & =\frac{\kappa_{\;\tau}^{\tau}}{4\pi} & =\frac{1}{4\pi}\left[\left\{ \frac{\ddot{a}+\frac{(M_{\theta+}a)-Q_{\theta+}^{2}}{a^{3}}}{\sqrt{1-\frac{2M_{\theta+}}{a}+\frac{Q_{\theta+}^{2}}{a^{2}}+\dot{a}^{2}}}\right\} -\left\{ \frac{\left(\ddot{a}-\frac{a}{\alpha^{2}}\right)}{\sqrt{\left(1-\frac{a^{2}}{\alpha^{2}}\right)+\dot{a}^{2}}}\right\} \right].\nonumber \\
\label{s2P}
\end{eqnarray}
\end{widetext}

To calculate the surface mass of the thin-shell, one can use this equation $M_{s}(a)=4\pi a^{2}\sigma$. To find stable solution, we consider a static case [$a_{0}\in(r_{-},r_{+})$].
\begin{widetext}
Then the surface charge and pressure at static case reduce to  
\begin{eqnarray}
\sigma(a_{0}) & = & -\frac{1}{4\pi a_{0}}\left[\sqrt{1-\frac{2M_{\theta+}}{a_{0}}+\frac{Q_{\theta+}^{2}}{a_{0}^{2}}}-\sqrt{\left(1-\frac{a_{0}^{2}}{\alpha^{2}}\right)}\right],\label{gen-surfenergy2a}\\
\P(a_{0}) & = & \frac{1}{8\pi a_{0}}\left[\frac{1-\frac{M_{\theta+}}{a_{0}}}{\sqrt{1-\frac{M_{\theta+}}{a_{0}}+\frac{Q_{\theta+}^{2}}{a_{0}^{2}}}}-\frac{\left(1-\frac{2a_{0}^2}{\alpha^{2}}\right)}{\sqrt{\left(1-\frac{a_{0}^{2}}{\alpha^{2}}\right)}}\right].\nonumber \\
\label{gen-surfpressure2a}
\end{eqnarray}
Then one can write that 
\begin{eqnarray}
\sigma(a_{0})+2\P(a_{0}) & = & \frac{1}{4\pi}\left[\left\{ \frac{\frac{(M_{\theta+}a_{0})-Q_{\theta+}^{2}}{a_{0}^{3}}}{\sqrt{1-\frac{2M_{\theta+}}{a_{0}}+\frac{Q_{\theta+}^{2}}{a_{0}^{2}}}}\right\} +\left\{ \frac{\left(\frac{a_{0}}{\alpha^{2}}\right)}{\sqrt{\left(1-\frac{a_{0}^{2}}{\alpha^{2}}\right)}}\right\} \right].\nonumber \\
\nonumber \\
\end{eqnarray}
\end{widetext}

Then we derive the conservation equation as follows: \begin{equation}
\frac{d(\sigma A)}{d\tau}+{\cal P}\,\frac{dA}{d\tau}=0\,.\label{E:conservation2}
\end{equation}
using the $S_{\;j|i}^{i}=\left[T_{\mu\nu}\;e_{\;(j)}^{\mu}n^{\nu}\right]_{-}^{+}$, where the surface area is $A=4\pi a^{2}$. 
One can write them also as follows: 
$\sigma'=-2\,(\sigma+{\cal P})/a$ ,
where $\sigma'=d\sigma/da$.

\section{Stability of the Charged Thin-shell Gravastars in Noncommutative Geometry}
In this section, we check the stability of the charged thin-shell gravastars in noncommutative geometry. To this purpose, we use the surface energy density  $\sigma(a)$  on the thin-shell of the gravastars as follows: 
\begin{equation} 
\frac{1}{2}\dot{a}^{2}+V(a)=0, 
\end{equation}
with the potential,
\begin{equation}
V(a)=\frac{1}{2}\left\{ 1-\frac{B(a)}{a}-\left[\frac{M_{s}(a)}{2a}\right]^{2}-\left[\frac{D(a)}{M_{s}(a)}\right]^{2}\right\} \,.\label{potential}
\end{equation}
It is noted that $B(a)$ and $D(a)$ are
\begin{widetext}
\begin{eqnarray}
B(a)=\frac{\left[\left(2M_{\theta+}-\frac{Q_{\theta+}^{2}}{a}\right)+(\frac{a^{3}}{\alpha^{2}})\right]}{2},\qquad D(a)=\left[\frac{\left[\left(2M_{\theta+}-\frac{Q_{\theta+}^{2}}{a}\right)-(\frac{a^{3}}{\alpha^{2}})\right]}{2}\right].\label{27}
\end{eqnarray}
One can also easily obtain the surface mass as a function of the potential:
\begin{equation}
M_{s}(a)=-a\left[\sqrt{1-\frac{2M_{\theta+}}{a}+\frac{Q_{\theta+}^{2}}{a^{2}}-2V(a)}-\sqrt{(1-\frac{a^{2}}{\alpha^{2}})-2V(a)}\right].\label{28}
\end{equation}
\end{widetext}
Then the surface charge and the pressure are rewritten in terms of potential as follows:
\begin{widetext}
\begin{equation}
\sigma=-\frac{1}{4\pi a}\left[\sqrt{1-\frac{2M_{\theta+}}{a}+\frac{Q_{\theta+}^{2}}{a^{2}}-2V}-\sqrt{(1-\frac{a^{2}}{\alpha^{2}})-2V}\right],
\end{equation}
\begin{equation}
{\cal P}=\frac{1}{8\pi a}\left[\frac{1-2V-aV'-\frac{M_{\theta+}}{a}}{\sqrt{1-\frac{2M_{\theta+}}{a}+\frac{Q_{\theta+}^{2}}{a^{2}}-2V}}-\frac{1-2V-aV'-\left(\frac{2a}{\alpha^{2}}\right)}{\sqrt{(1-\frac{a^{2}}{\alpha^{2}})-2V}}\right].\label{gen-surfpressure2-onshell}
\end{equation}
\end{widetext}

To find the stable solution, we linearize it using the Taylor expansion  around the $a_{0}$ to second order as follows:
\begin{equation}
V(a)=\frac{1}{2}V''(a_{0})(a-a_{0})^{2}+O[(a-a_{0})^{3}]\,.\label{linear-potential}
\end{equation}
Note that for stability, the conditions are  $V(a_{0})=V'(a_{0})=0$, $\dot{a}_{0}=\ddot{a}_{0}=0$
 and $V''(a_{0})>0$.
Using the relation $M_{s}(a)=4\pi\sigma(a)a^{2}$, we use the $M_{s}''(a_{0})$ instead of $V''(a_{0})\geq0$ as following \cite{Lobo:2015lbc}:
\begin{widetext}
\begin{equation}
M_{s}''(a_{0})\geq\frac{1}{4 a_{0}^{3}}\left\{ \frac{[\frac{2\left(M_{\theta+}+Q_{\theta+}^{2}\right)}{a_{0}}]^{2}}{[1-\frac{2M_{\theta+}}{a_{0}}+\frac{Q_{\theta+}^{2}}{a_{0}^{2}}]^{3/2}}-\frac{[\frac{-a_{0}^{3}}{\alpha^{2}}]^{2}}{[(1-\frac{a_{0}^{2}}{\alpha^{2}})]^{3/2}}\right\} +\frac{1}{2}\left\{ \frac{\frac{2Q_{\theta+}^{2}}{a_{0}^{3}}}{\sqrt{1-\frac{2M_{\theta+}}{a_{0}}+\frac{Q_{\theta+}^{2}}{a_{0}^{2}}}}-\frac{\frac{4a_{0}}{\alpha^{2}}}{\sqrt{(1-\frac{a_{0}^{2}}{\alpha^{2}})}}\right\} ,\label{stable_ddms1}
\end{equation}

so for the stable solution, it must satisfy the above relation as shown in Fig. (1). Note that we have used the following equation 
\begin{eqnarray}\notag
V''(a_{0})&=&-\frac{3 M_s^2(a_0)}{4 a_0^4}+\Big[\frac{M'_s(a_0)}{a_0^3}-\frac{M''_s(a_0)}{4 a_0^2}\Big]M_s(a_0)-\frac{B''(a_0)}{2 a_0}+\frac{B'(a_0)}{a_0^2}-\frac{B(a_0)}{a_0^3}-\frac{M'_s(a_0)^2}{4 a_0^2}\\
&-&\frac{D'^2(a_{0})+D(a_{0}) D''(a_{0})}{M_s^2(a_{0})}+\frac{4 D(a_{0}) D'(a_{0}) M'_s(a_{0})+D^2(a_{0}) M''_s(a_{0})}{M_s^3(a_{0})}-\frac{3 D^2(a_{0}) (M'_s)^2(a_{0})}{M_s^4(a_{0})},
\end{eqnarray}
\end{widetext}
where
\begin{equation}
M'_s(a_{0})=8 \pi a_0 \sigma_0-8 \pi a_0 (\sigma_0+p_0),
\end{equation}
and
\begin{eqnarray}\notag
M''_s(a_{0})&=& 8 \pi  \sigma_0-32 \pi(\sigma_0+p_0)\\
& + & 4\pi \left[ 2(\sigma_0+p_0)+4 (\sigma_0+p_0)(1+\eta)\right].
\end{eqnarray}

Moreover, we have also introduced $\eta(a)=\mathcal{P}'(a)/\sigma'(a)|_{a_0}$, as a parameter 
which will play a fundamental role in determining the stability regions of the respective solutions.
Generally, $\eta$ interpreted as the speed of sound, so that one would expect the range of $0 < \eta \leq 1$,
that the speed of sound should not exceed the speed of light. But the range of ${\eta}$ may be 
lying outside the range of $0 < \eta \leq 1$, on the surface layer and for extensive discussion
see Refs. \cite{Poisson}. Therefore, in this work the range of $\eta$  will be relaxed and we use
graphical reputation to determine the stability regions given by the Eq. (34), due to the 
complexity of the expression.  

\begin{figure}[h!]
\includegraphics[width=0.33\textwidth]{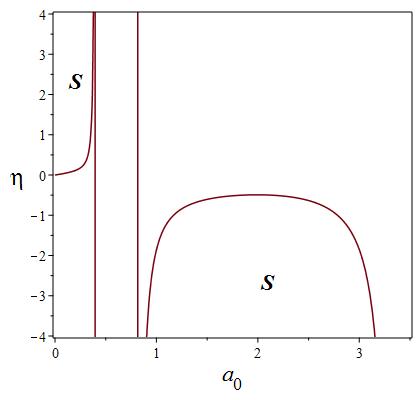} %
\caption{\small \textit{Stability regions of the charged gravastar in terms of $\eta=P'/\sigma '$ as a function of $a_{0}$. We choose $M_{\theta}=2$, $Q_{\theta}=1.5$, $\alpha=0.4$.} }
\end{figure}

\begin{figure}[h!]
\includegraphics[width=0.33\textwidth]{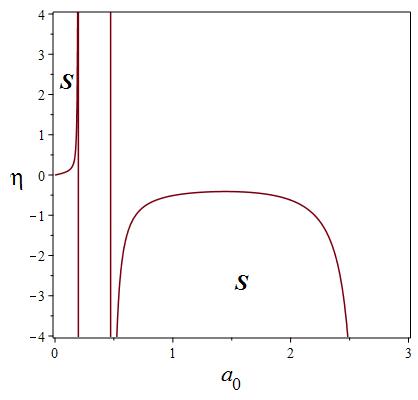} %
\caption{\small \textit{Stability regions of the charged gravastar in terms of $\eta=P'/\sigma '$ as a function of $a_{0}$. We choose $M_{\theta}=1.5$, $Q_{\theta}=1$, $\alpha=0.2$.} }
\end{figure}

\begin{figure}[h!]
\includegraphics[width=0.33\textwidth]{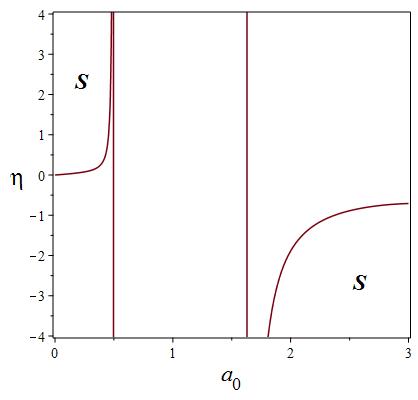} %
\caption{\small \textit{Stability regions of the charged gravastar in terms of $\eta=P'/\sigma '$ as a function of $a_{0}$. We choose $M_{\theta}=3$, $Q_{\theta}=2.5$, $\alpha=0.5$.} }
\end{figure}

\section{Thermodynamics and stability conditions for the thin shell}

Now, we turn to the thermodynamical stability of the thin-shell.
Following \cite{Lemos}, we assume that the shell is in thermal equilibrium,
with a locally measured temperature $T$ and an entropy $S$.  Here the entropy $S$
can be expressed as a function of the state independent variables of
surface mass of the thin shell $M$, area $A$, and charge $Q$. Thus the first law of
thermodynamics provides the following relationship
\begin{equation}
T dS = dM +pdA- \Phi dQ,
\end{equation}
where $\left(M, A, Q\right)$ can be considered as three generic parameters.
It is important to note that we consider the particles N is constant. Now it is
a simple matter, to obtain the entropy S, we shall adopt three equations of state,namely,
$p\left(M, A, Q\right)$, $\beta \left(M, A, Q\right)$, and $\Phi \left(M, A, Q\right)$
namely, the pressure, temperature, and charge equations of state, respectively
and we define the inverse temperature $\beta \equiv 1/T$.

It is of particular interest to obtain an expression for the entropy, the integrability
conditions must be specified, which follow directly from the first law of thermodynamics are given by

\begin{equation}
\left( \frac{\partial \beta}{\partial A} \right)= \left( \frac{\partial \beta p}{\partial M} \right)_{A,Q},
\end{equation}
\begin{equation}
\left( \frac{\partial \beta}{\partial Q} \right)= \left( \frac{\partial \beta \Phi}{\partial M} \right)_{A,Q},
\end{equation}
\begin{equation}
\left( \frac{\partial \beta p}{\partial Q} \right)= \left( \frac{\partial \beta \Phi}{\partial A} \right)_{M,Q}.
\end{equation}

Thus, one may easily determine the relations between the three EOS of the system.
This result also originates for studying the local intrinsic stability of the shell, by the 
first law in Eq. (2). It is more convenient to work out the thermodynamic stability 
are dictated by the following inequalities :
\begin{equation}
\left( \frac{\partial^2 S}{\partial M^2} \right)_{A,Q} \leq 0,
\end{equation}

\begin{equation}
\left( \frac{\partial^2 S}{\partial A^2} \right)_{M,Q} \leq 0,
\end{equation}

\begin{equation}
\left( \frac{\partial^2 S}{\partial Q^2} \right)_{M,A} \leq 0,
\end{equation}

\begin{equation}
\left( \frac{\partial^2 S}{\partial M^2} \right)\left( \frac{\partial^2 S}{\partial A^2} \right)-
\left( \frac{\partial^2 S}{\partial M \partial A} \right)^{2} \geq 0,
\end{equation}

\begin{equation}
\left( \frac{\partial^2 S}{\partial A^2} \right)\left( \frac{\partial^2 S}{\partial Q^2} \right)-
\left( \frac{\partial^2 S}{\partial A \partial Q} \right)^{2} \geq 0,
\end{equation}

\begin{equation}
\left( \frac{\partial^2 S}{\partial M^2} \right)\left( \frac{\partial^2 S}{\partial Q^2} \right)-
\left( \frac{\partial^2 S}{\partial M \partial Q} \right)^{2} \geq 0,
\end{equation}

\begin{equation}
\left( \frac{\partial^2 S}{\partial M^2} \right)\left( \frac{\partial^2 S}{\partial Q \partial A} \right)-
\left( \frac{\partial^2 S}{\partial M \partial A} \right)
\left( \frac{\partial^2 S}{\partial M \partial Q} \right) \geq 0,
\end{equation}
For more discussion and derivation of these expression see Refs. \cite{Quinta}.

\section{Conclusions}

In this paper, we have studied the stability of a particular class of
thin-shell gravastar solutions, in the context of charged noncommutative geometry.
For this purpose we consider the de Sitter geometry in the interior
of the gravastar by matching an exterior charged noncommutative solution at a junction interface situated outside the event horizon. We showed that gravastar's 
shell satisfies the null energy conditions in Fig. (1).

We further explored the gravastar solution by the dynamical stability of the
transition layer, which is sufficient close to the event horizon.
It is found that for specific choices of mass $M_{\theta}$, charge $Q_{\theta}$ and the values of $\alpha$, the stable configurations of the surface layer do exists
which is sufficiently close to where the event horizon is expected to form.
In next we explore the thermodynamical stability of the thin-shell gravastar, 
using the shell in the thermal equilibrium.
\begin{acknowledgments}
This work was supported by the Chilean FONDECYT Grant No. 3170035 (A\"{O}).
\end{acknowledgments}

\end{document}